\documentclass[aps,prb,superscriptaddress,citeautoscript,twocolumn,longbibliography,showkeys]{revtex4-1}
\usepackage{graphicx}
\usepackage{amsfonts}
\usepackage{amsmath}
\usepackage{amssymb}
\usepackage{bm}
\usepackage{textcomp}
\usepackage{multirow}

\usepackage{bbold}
\usepackage{mathtools}
\usepackage{dcolumn}
\usepackage{hyperref}
\usepackage[usenames]{color}
\usepackage{subcaption}

\usepackage{physics}
\usepackage{units}

\usepackage{tablefootnote}
\usepackage{natbib}

\usepackage{xcolor}

\AtBeginDocument{%
	\newwrite\bibnotes
	\def\bibnotesext{Notes.bib}
	\immediate\openout\bibnotes=\jobname\bibnotesext
	\immediate\write\bibnotes{@CONTROL{REVTEX41Control}}
	\immediate\write\bibnotes{@CONTROL{%
			apsrev41Control,author="08",editor="1",pages="1",title="0",year="1"}}
	\if@filesw
	\immediate\write\@auxout{\string\citation{apsrev41Control}}%
	\fi
}%

\usepackage{ragged2e}
\DeclareCaptionJustification{justified}{\justifying}
\def\beq{\begin{equation}}
\def\eeq{\end{equation}}
\def\vk{{\bf k}}
\def\vq{{\bf q}}
\def\vR{{\bf R}}

\newcommand{\out}[1]{{}}

\begin{document}

%\title{Structure of the hidden order and multipolar exchange springs in NpO$_{2}$}
\title{Hidden order and multipolar exchange striction in a correlated $f$-electron system}

\author{Leonid V. Pourovskii}
\address{Centre de Physique Th\'eorique, Ecole Polytechnique, CNRS, Institut Polytechnique de Paris, 91128 Palaiseau Cedex, France}
\address{Coll\`ege de France, 11 place Marcelin Berthelot, 75005 Paris, France}

\author{Sergii Khmelevskyi}
\address{Research Center for Computational Materials Science and Engineering, Vienna University of Technology, Karlsplatz 13,  1040 Vienna, Austria}

\begin{abstract}
	\vspace{5mm}
	\begin{center} {\bf Abstract} \end{center}
The nature of order 
%parameter
 in low-temperature phases of some materials  is not directly seen by experiment.
 Such ``hidden orders'' (HO)  may inspire decades of research to identify the mechanism underlying  
those exotic states of matter. 
In insulators, HO phases originate in degenerate many-electron states on localized $f$ or $d$ shells that may  harbor high-rank multipole moments. Coupled by inter-site exchange, those moments form a vast space of competing order parameters.  
%The dioxide of neptunium NpO$_2$ is a 
%oldest known 
%prototypical HO system, where the primary order parameter is believed to be of  rank-5 (triakontadipolar), 
%as has been inferred from various experimental probes. %
%but the  microscopic interactions driving this order are still not resolved. 
Here, we show 
%how a low-energy Hamiltonian  of NpO$_2$ is derived by a many-body {\it ab initio} force-theorem method from its symmetry-unbroken paramagnetic phase. 
how the ground state order and magnetic excitations of a prototypical HO system, neptunium dioxide   NpO$_2$, can be fully described by a low-energy Hamiltonian derived by a many-body {\it ab initio} force-theorem method. 
Superexchange interactions between the  lowest 
%$\Gamma_{8}$
 crystal-field  quadruplet of Np$^{4+}$ ions  induce a 
 %3$\vk$ antiferromagnetic (AF) order of  pseudo-octupole and pseudo-dipole $\Gamma_{8}$ moments. Transformed to observable multipolar moments this  corresponds to a 
 primary non-collinear order of  time-odd rank-5 (triakontadipolar) moments  with a secondary %longitudinal 
  quadrupole order preserving the cubic symmetry of NpO$_2$.  
  %The calculated on-site exchange splitting  and magnetic excitation spectra of the HO phase agree well with experiment. 
  % An observed  anomalous volume contraction in the NpO$_2$ HO phase  is shown to be induced by a two-site multipolar exchange striction.
  Our study also reveals an unconventional multipolar exchange-striction mechanism behind the anomalous volume contraction of the NpO$_2$ HO phase.
  	\vspace{10mm}
  \begin{center} {\bf Significance statement} \end{center}
Second-order phase transitions in solids occur due to spontaneous symmetry breaking with an order parameter continuously emerging from the disordered high-temperature phase. In some materials, the phase transitions is clearly detected in thermodynamic functions (e.g., specific heat) but the microscopic order parameters remain  “hidden” from researchers, in some cases for decades. Here we show how such hidden-order parameters can be unambiguously  identified and  the corresponding ordered phase fully described using a first-principles many-body linear-response theory. Considering  the canonical “hidden-order" system neptunium dioxide,  we also identify a novel mechanism of spontaneous multipolar exchange striction that induces an anomalous volume contraction of the  hidden-order phase in NpO$_2$. 
\end{abstract}

\keywords{strongly correlated electrons; hidden order; ab initio; linear response; effective Hamiltonians}

\maketitle
%\date{\today}

%\section*{INTRODUCTION}

Spontaneous symmetry breaking, or a phase transition, in extended
systems is associated with emergence of a macroscopic order parameter, 
which is the statistical average over some physical observable. 
%In
%solids breaking of the time-reversal symmetry  results in a magnetic
%order and superconductivity originates from breaking of the gauge invariance.
% In some systems  the nature of order parameter is not directly seen by experiment. Such ``hidden orders'' (HO)  may inspire decades of research to identify the mechanism underlying  
In some systems the onset of order is clearly observed in the behavior of thermodynamic functions, however,  the order parameter is not detectable by standard probes like neutron scattering or magnetic susceptibility measurements.  In metals such phenomena are typically associated with strongly correlated heavy-fermion behavior, as in the case of enigmatic  URu$_2$Si$_2$ 
 \cite{Haule2009,OppeneerRMP,Coleman2013}. 
In correlated insulators
 HO phases typically originate in high-rank multipolar  degrees of freedom on localized $f$ and $d$-electron shells. 
It has been realized long time ago \cite{Baker1971} that in correlated magnetic insulators with strong spin-orbit coupling, apart from the ordinary Heisenberg  interaction between
localized spins,  there may also exist inter-site interactions
coupling higher order spin operators (magnetic multipoles).
If those interactions are essentially large this might lead to a
new state of matter - a multipolar order without any associated magnetic order\cite{ChenLexy1971,AndeevJETP,ChandreColemannematics}.
%In the simplest case it is a nematic order of the magnetic quadrupoles due to bi-quadratic exchange coupling. 
%However,  the majority of correlated insulators with strong spin-orbit coupling do exhibit the usual  magnetic order, with one-site multipolar moments being  
%``slaved'' to dipolar ones  and manifesting themselves, e.~g., via lattice distortions in the magnetically ordered phase \cite{Morin}. 
%phase due to magneto-elastic coupling which connects deformation energy
%with magnetic quadrupoles \cite{Morin}. 
%So far a purely multipolar order due to high-rank multipoles has been suggested only for few systems \cite{Santini2009,OpennerJPJ,Morin}, each such case attracting considerable interest. 
Such purely multipolar orders have been observed in various $f$-electron\cite{OppeneerRMP,Santini2009,Cameron2016,Sato2012,Tsujimoto2014} and  transition-metal correlated systems \cite{Lu2017,Hirai2019,Maharaj2020}. 
Owing to numerous competing order parameters and small relevant energy scales, 
 quantitative description of the HO represents a 
 formidable challenge for theory. First-principles approaches to HO are typically restricted to simulations of few likely phases inferred experimentally and  do not attempt to explore the full space of possible multipolar orders\cite{Haule2009,Suzuki_oppeneerAnO22013}.  

 Difficulties of identifying the physical order parameter in a vast phase space of possible HOs are  exemplified by the case of cubic NpO$_2$\cite{Santini2009}. %This system is the oldest known candidate for a purely high-rank magnetic multipolar  
%order (MMO) .
A sharp second order 
phase transition at $T_0\simeq$26~K  was detected in NpO$_2$ more than half-century ago \cite{Osborn53},
with no evidence for underlying magnetic order and structural transformations
\cite{DUNLAP19681365,COX19671649,Amoretti_1992}, apart from
%As
%the understanding of the nematic order concept has progressed over
%time \cite{ChenLexy1971} 
a small anomalous contraction of the cubic unit cell volume observed\cite{Lander_nodistrotionNpO2} 
below $T_0$. At the same time, NMR measurements\cite{127_tokunaga} detect two inequivalent oxygen sites  in the unit cell below $T_0$, due to  lowering of the cubic symmetry from high-temperature
$Fm\bar{3}m$ to $Pn\bar{3}m$ by a longitudinal order of Np  quadrupoles. 
However, a  primary quadrupolar order parameter initially suggested\cite{ERDOS1980164} is excluded since muon spin-rotation measurements detect a non-zero magnetic density\cite{KOPMANN1998463}. Moreover, the  crystal-field (CF) ground state quadruplet $\Gamma_8$ is split in the HO phase  into two singlets and a doublet suggesting a time-odd primary order parameter\cite{Amoretti_1992,136_Santini_Garreta2006,Magnani2008}. There is a multitude of possible high-rank odd order parameters  realizable within the $J=$9/2 ground state multiplet of Np$^{4+}$ ($J=$3,5,7, i.e. octupolar, triakontadipolar, etc.).  A lot of efforts has been
directed  over
the last two decades \cite{134_Tokunaga2006,127_tokunaga,135_Sakaj_JPSJ2005,136_Santini_Garreta2006,137_Kubo2005j-j,NMR2019,KOPMANN1998463} to identify a  possible primary order able to reconcile various experimental facts.  In particular, assumption of 
a triakontadipolar AF 3\vk\  order of the $\Gamma_5$ symmetry was shown\cite{Paixao2002,136_Santini_Garreta2006,Magnani2008}  to lead to the best agreement with X-ray scattering and inelastic neutron scattering (INS) spectra. The same hypothesis is also strongly supported by estimates of the relative strength of odd $\Gamma_5$ multipole moments  on the Np $f^3$ shell in the presence of CF splitting\cite{Magnani2005,136_Santini_Garreta2006}.
% , which is predicted to suppress octupolar (as well as rank-7) moments and to enhance  trikontadipolar ones.
%for the explanation of the enigmatic transition in NpO\textsubscript{2}

Though there are substantial experimental evidences to support the 3\vk\ rank-5 order  in NpO$_2$, 
%is quite convincingly established by experimen,
 the mechanism of its formation is still not well understood. The simplest possible form of the superexchange (SE) Hamiltonian, consisting of  diagonal nearest-neighbor interactions between three $\Gamma_5$ triakontadipoles and between three dipole moments, has been employed in analysis of low-temperature susceptibility and INS data\cite{136_Santini_Garreta2006,Magnani2008}. The full structure of this Hamiltonian cannot be extracted from experiment due to a large number of possible SE interactions (SEI). The measured CF splitting of 55~meV between the $\Gamma_8$ ground-state  and first exited CF level\cite{Amoretti_1992,Fournier1991,Santini2009} is much larger than $T_0$, suggesting SEI between $\Gamma_8$ states on neighboring Np ions as the origin of its exotic ordered phase.  Such low-energy SE Hamiltonian has not been so far derived theoretically. Previous theoretical density-functional-theory+U (DFT+U)  studies 
% taking into the account the correlated character of the 5f-shell by including U parameter~\cite{Anisimov1997} 
have confirmed the stability of a triakontadipolar order in NpO$_2$\cite{Suzuki_oppeneer2010npo2,Suzuki_oppeneerAnO22013}, however, they
 imposed an initial symmetry breaking consistent with the 3\vk\ rank-5 order inferred experimentally.
%, hence, possible competing orders were not treated on the equal footing. 

In this work we apply a novel framework to the problem of "hidden" multipolar orders in correlated insulators as exemplified by NpO$_2$. It consists in evaluating the full low-energy SE Hamiltonian from an {\it ab initio} description of the symmetric paramagnetic phase.    We start with charge-self consistent  DFT+dynamical mean-field theory\cite{Georges1996,Anisimov1997_1,Aichhorn2016} calculations for paramagnetic NpO$_2$ treating Np 5$f$ within the quasi-atomic Hubbard-I (HI) approximation\cite{hubbard_1} (this method is abbreviated below as DFT+HI) to obtain its electronic structure and the composition of a CF-split Np 5$f^3$  shell. A  force theorem approach\cite{Pourovskii2016}, abbreviated FT-HI, is subsequently employed to derive SEIs between  the calculated CF $\Gamma_8$  ground-state quadruplets. Our study represents first, to our awareness, {\it ab initio} electronic structure calculation  of a complete  SE Hamiltonian for high-rank magnetic multipoles in  an $f$-electron crystalline material.    By solving this Hamiltonian, we find a 3\vk\   rank-5 primary magnetic multipolar order (MMO) accompanied by a secondary longitudinal  quarupole order.  The calculated time-odd  splitting of $\Gamma_8$   and  magnetic excitation spectra are in a good agreement with experiment. The lattice contraction effect\cite{Lander_nodistrotionNpO2} 
%due to the MMO in NpO$_2$ is shown to stem from a volume dependence of the calculated SEI, which describes  this  small peculiarity with a quantitative precision.   
is shown to be not related to quadrupole ordering as suggested before\cite{2003JPCM_magnetovolume}, but rather to stem from the volume dependence of leading time-odd SEIs. This multipolar "exchange-spring" effect is quite unique and has not been, to our awareness, discussed previously in the literature.  
Overall, we show that within our first-principles scheme, which treats all competing order parameters on the equal footing,  high-rank multipolar orders in correlated insulators can be captured both qualitatively and quantitatively. 

\section*{Results}

\subsection*{Crystal-field splitting and super-exchange Hamiltonian}

\begin{figure*}[!tp]
	\begin{centering}
		\includegraphics[width=1.9\columnwidth]{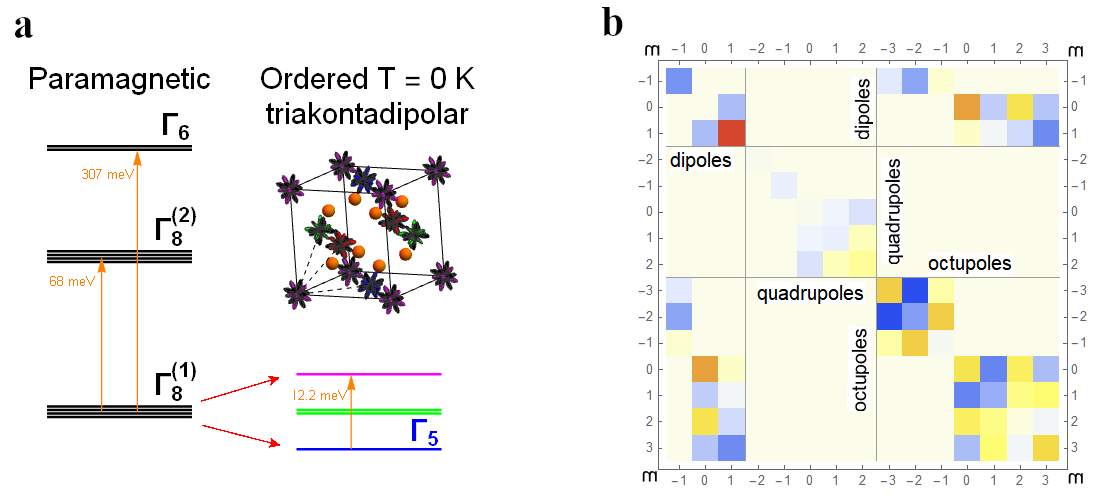} 
		\par\end{centering}
	\caption{{\bf Np 5$f$ on-site splitting and Np-Np inter-site interactions in NpO$_2$.} {\bf (a)} Calculated crystal-field (CF) spiting of Np $5f^3$ J=9/2 multiplet 
		in paramagnetic state (left) and exchange splitting of the ground state
		quartet in the predicted ordered phase at zero temperature (right). Note the energy
		rescaling for the $\Gamma_6$ level on the CF splitting plot. 
		%The percentages above the 
		%red arrows indicate the admixture of the corresponding CF level  
		%into the ordered $\Gamma_5$ ground state induced by super-exchange interactions (SEI).
		Inset: the crystal structure of the NpO$_2$. Positions of the Np-atoms is surrounded
		by the polar plot of the calculated primary order parameter (triakontadipoles)
		in the ground state. Different colors indicate four non-equivalent Np
		positions in the ordered state. The orange spheres are oxygen atoms. {\bf (b)} SEI matrix between nearest-neighbor Np in NpO$_2$.  These SEI couple multipolar operators 
		$\hat{O}_{lm}$ in the $J_{eff}=$3/2 space of the $\Gamma_{8}$  CF ground state.  Their values in the local coordinate
		system (with z-axis directed along the given bond direction and y-axis
		along the edge of the fcc lattice)  are presented as a temperature map with the warm and cool colors representing, respectively,  antiferromagnetic  and  ferromagnetic coupling of the corresponding multipoles.% \lp{Scale should be added to the colormap}. 
		%Full definitions of multipoles
		%Olm operators and the data in the form of the numerical table see
		%Supplement.
	}	
	\label{fig1} 
\end{figure*}

We start with evaluating the Np  CF  splitting  in paramagnetic NpO$_2$; as discussed above, this splitting determines the space of low-energy states forming the MMO. The  CF splitting of the Np $5f^3$  ground state multiplet $J=$9/2 calculated by DFT+HI is shown
in Fig.\ref{fig1}a. The ground state $\Gamma_{8}$
quartet is separated from another, excited, $\Gamma_{8}$ quartet by 68 meV, 
in agreement with the experimental range for this splitting, 30-80 meV, inferred from  INS mesuarements
 \cite{CACIUFFO1991197,Amoretti_1992,Fournier1991}. The
broad experimental range for the excited $\Gamma_8$ energy is due  to the presence of a dispersive phonon
branch in the same range (this overlap has been a major source of difficulties for the phenomenological
analysis of the NpO$_2$ in the framework of crystalline
effective filed  model \cite{Santini2009}). Third exited level,
 $\Gamma_{6}$ doublet, is located  at much higher energy above 300~meV.

 Our calculated CF corresponds to $x=$-0.54 parameterizing the relative magnitude of the 4 and 6-order contributions to the cubic CF \cite{Lea1962}. Our values are in good agreement with $x=$-0.48 inferred from INS measurements\cite{Amoretti_1992} and  analysis of CF excitation energies along the actinide dioxide series\cite{Magnani2005}. The  CF level energies calculated within DFT+HI are also in good agreement with previous DFT+DMFT calculations by Koloren\v{c} {\it et al.}\cite{Kolorenc2015} employing an exact diagonalization approach to Np 5$f$. The calculated wavefunctions of the CF ground-state $\Gamma_8$ quartet (see Supplementary Table~I) feature a small admixture  of the excited $J=$11/2 and 13/2 multiplets.
%The calculated eigenstates of the ground state quartet in the $|J;m_{J}\rangle$
%basis is of the form 
%\begin{flalign}
%|3/2\rangle= & 0.908|4;+3\rangle-0.343|4;-1\rangle-0.032|5;-5\rangle+\label{Gamma5_WF}\\
%|1/2\rangle= & 0.686|4;+2\rangle-0.686|4;-2\rangle-0.033|5;-2\rangle+\nonumber \\
%\nonumber \\
%\hline |-1/2\rangle= & -0.908|4;-3\rangle+0.343|4;+1\rangle-0.032|5;+5\rangle\nonumber 
%\end{flalign}
%The  possible impact of $J-J$ mixing on the MMO in NpO\textsubscript{2} has
%been previously  discussed on a phenomenological basis \cite{Magnani2005,137_Kubo2005j-j}.
%However, the  predicted $J=$11/2 admixture  into the $\Gamma_8$ ground
%state appears to be too small ($\sim$2\%) to significantly
%alter conclusions  derived from the bare LS coupling scheme.
%The phase factors in the equ.\ref{Gamma5_WF} has been chosen in way
%that presented there functions satisfying symmetry conditions \cite{Chibotaru2018}
%to serve as a basis for the true 3/2-pseudospin space. 

The space of CF GS $\Gamma_8$ quadruplet  is conveniently represented by  the effective angular momentum  quantum number $J_
{eff}$=3/2, with the $\Gamma_8$  wave functions (WFs) labeled by the corresponding projection $M=-3/2...+3/2$ and the phases of WFs chosen to satisfy the time-reversal symmetry (see Supplementary Table~I).  The on-site degrees of freedom within the GS quadruplet are (pseudo) dipole, quadrupole, and octupole moments defined for  $J_{eff}$=3/2 in the standard way\cite{Santini2009}. Hence, the most general form for a SE coupling between $\Gamma_8$ quadruplets  on two Np sites  reads
\begin{flalign}\label{eq:SE_bond}
\sum_{KQK'Q'}V_{KK'}^{QQ'}(\vR)O_K^Q(\vR_0)O_{K'}^{Q'}(\vR_0+\vR),
\end{flalign}
where $O_K^Q(\vR_0)$ and $O_K^Q(\vR_0+\vR)$ are the real spherical tensor operators\cite{Santini2009} for $J=$3/2 of rank $K=$1, 2, or 3,  $-K \le Q \le K$, acting on the site $\vR_0$ and $\vR_0+\vR$, respectively, $V_{KK'}^{QQ'}(\vR)$ is the  SEI that couples them; due to the translational invariance, $V_{KK'}^{QQ'}$ depends only on the intersite lattice vector $\vR$.  

We employed the FT-HI approach~\cite{Pourovskii2016} to evaluate all interactions $V_{KK'}^{QQ'}(\vR)$ from the converged DFT+HubI NpO$_2$ electronic structure for several first Np coordination shells.  
Only nearest-neighbor (NN) SEIs are significant, with longer distance ones being more than an order of magnitude smaller. 
All SEIs $V_{KK'}^{QQ'}(\vR)$ for a given bond $\vR$ form a matrix, designated below as the SEI matrix, with the rows and columns labeled by the moments $KQ$ and $K'Q'$ on the sites $\vR_0$ and $\vR_0+\vR$, respectively. 
 The NN SEI matrix $\hat{V}(\vR)$ is graphically represented  in Fig.~\ref{fig1}b  (see also Supplementary Table~II) using  a local coordinate system with the quantization axis $z||\vR$ .  This $15\times 15$ matrix is of a block-diagonal form, since the interactions between time-even and time-odd moments  
%(dipole-quadrupole and quadrupole-octupole) 
are zero by symmetry. It can thus be separated into  
% $\hat{V}_{DD}$, $\hat{V}_{QQ}$, $\hat{V}_{OO}$, $\hat{H}_{DO}$ blocks, namely,
 the dipole-dipole (DD), quadrupole-quadrupole (QQ), octupole-octupole (OO), and dipole-octupole (DO) blocks. In spite of this simplification, the SEI matrix $\hat{V}$ can in principle contain 70 distinct elements. The number of distinct non-zero matrix elements in $\hat{V}$, while reduced by the cubic symmetry to 38, remains rather large.

Our calculations predict the largest values for the diagonal DD $x-x$ (AF, 1.6~meV) SEI.  However, the off-diagonal  OO $xyz$ to $y(x^2-3y^3)$ (ferro, -1.5 meV)  and DO $z$ to $z^3$ (AF, 0.95~meV) couplings are of about the same magnitude as the  DD $x-x$ one. Overall, the calculated $\hat{V}$ matrix shown in Fig.~\ref{fig1}b features many non-negligible DD, OO, and DO interactions of a comparable magnitude. The QQ interactions are weaker reflecting the secondary nature of the quadupole order. Our calculations thus predict a complex and frustrated SE in NpO$_2$,  which may give rise to multiple competing time-odd orders. 
 Therefore, as has been previously noted \cite{Magnani2008,Santini2009},  extracting a full SE Hamiltonian of NpO$_2$ from experimental (e.~g., INS) data is virtually impossible due to a large number of parameters entering into the fit of an excitation spectra. The same difficulty is encountered by {\it ab initio} approaches based on total energy calculations for symmetry broken phases\cite{Suzuki_oppeneer2010npo2,Suzuki_oppeneerAnO22013,Pi2014},  which require a large number  of very precises calculations to extract multiple non-negligible matrix elements of $\hat{V}$, with the magnitude of 0.5~meV and above.  Within the present framework all interactions are extracted from a single {\it ab initio} calculation for paramagnetic NpO$_2$.
%} 

\subsection*{Ordered state of NpO$_2$}

The calculated SE Hamiltonian for NpO$_2$ reads
\beq\label{HSE}
H_{SE}=\frac{1}{N}\sum_{\vR \in \mathrm{NN}}\hat{O}(\vR_0) \hat{V}(\vR)\hat{O}(\vR_0+\vR),
\eeq
where $\vR$ runs over all NN bonds in the Np fcc sublattice, $N$ is the number of Np sites, and we also introduced  the obvious vector notation $\hat{O} \equiv [O_1^{-1},..O_3^3]$ for multipole tensors. The SEI  matrix $\hat{V}(\vR)$, eq.~\ref{eq:SE_bond}, is  that in the local frame $\vR ||z$  (Fig.~\ref{fig1}b) rotated to align $\vR$ along the corresponding Np NN bond. We solved (\ref{HSE}) numerically\cite{} within the mean-field (MF) approximation, obtaining a second-order transition at $T_0=$38~K, in good agreement with its experimental value of 26~K taking into account the usual mean-field overestimation of ordering temperatures. The numerical results were verified by a linearized (MF) theory,  derived by first-order expansion of MF equations in the order parameters $\langle \hat{O}\rangle$, see Method section. 

The resulting GS order of NpO$_2$ in the $J_{eff}$ space consists of a primary (pseudo) dipole-octupole order combined with secondary (pseudo) quadrupole one (see Supplementary Table~III  for the values of all $J_{eff}$ order parameters and  Supplementary Figure~1). The pseudo-dipole order  is a complex 3\vk\ planar  AF structure, 
with four inequivalent simple-cubic sublattices forming two pairs with different moment magnitude; the moments of those two pairs are aligned
along the $\langle1,1,0\rangle$ and $\langle3,1,0\rangle$ directions in the fcc lattice, respectively. The origin of this  inclined AF structure of pseudo-dipoles is in the dipole-octupole SEI; similarly to physical AF magnetic orders found in some materials with large SO that are likely induced by
higher-order magnetic multipolar interactions. The pseudo-octupoles order is also oriented in non-symmetrical
directions. The secondary psedo-quadrupole order is of a 3\vk\    type, which we analyze in details below.

\begin{figure*}[!tb]
	\begin{centering}
		\includegraphics[width=2.0\columnwidth]{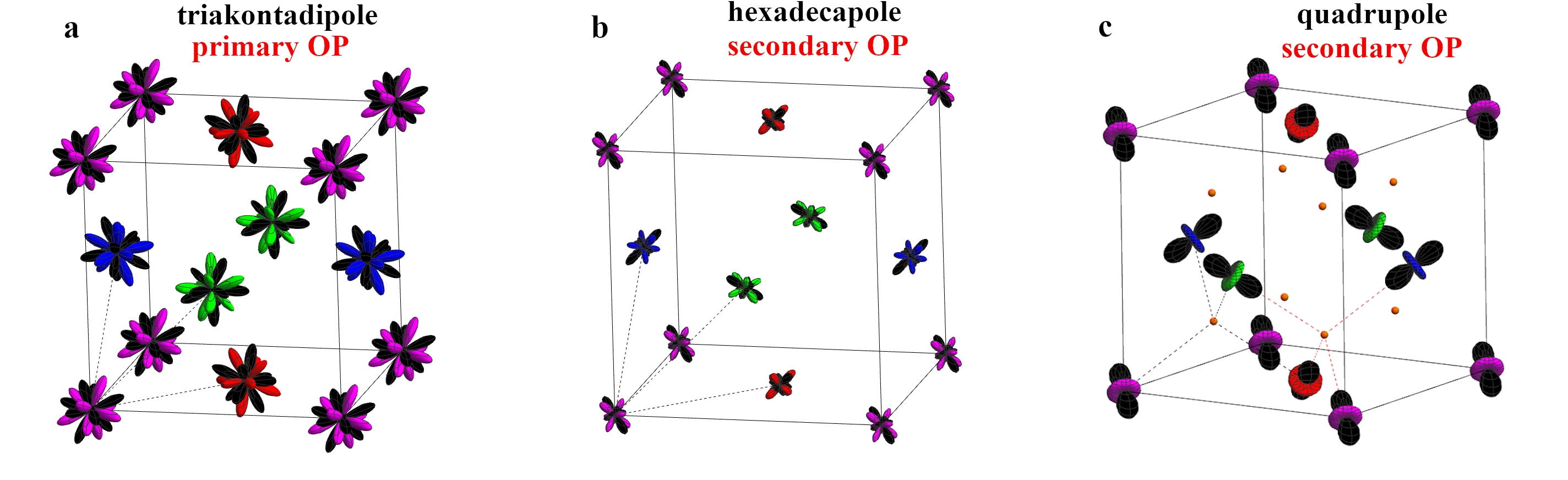} 
		\par\end{centering}
	\caption{{\bf Calculated multipolar order in NpO$_2$.}
		This order of the physical multipoles in NpO$_2$
		derived by mapping of the $J _{eff}=$3/2 space to the full $J=$9/2 
		space. The physical magnetic dipoles (magnetic
		moments) are exactly canceled on all Np sites, resulting in a purely
		multipolar order. 	{\bf (a)} The primary physical order parameter, $K=$5 triakontadipole
		({\bf b}) The largest secondary ``slave'' order parameter, $K=$4 hexadecapole. The displayed isovalue  for the primary and
		secondary order parameters
		is normalized with respect to the maximum possible value of a given moment
		in the $\Gamma_8$ crystal-field ground-state quartet. Hence the relative size of plotted moments indicates their relative magnitude. Black dashed lines are the primitive lattice translations of the original Np cubic face-centered sublattice, which in the ordered phase connect inequivalent Np sublattices.   ({\bf  c})   Longitudinal 3\vk\  order of  ``slave'' quadrupole moments, which  are scaled on the plot by a factor of 4.  Np-O bonds for two inequivalent O sites  (small orange balls) in ordered NpO$_2$ are indicated by dashed lines. One sees that in the Np tetrahedron around each O all ordered Np quadrupoles are directed either along the Np-O bonds, or form the same angle with  those bonds. 
		This longitudinal  3\vk\ quadrupolar order preserves  the cubic positions of oxygen sites in the ordered phase of NpO$_2$, in contrast to  the transverse 3\vk\  quadrupole order in UO$_2$. }
	
	\label{fig_2} 
\end{figure*}

We subsequently mapped the moments calculated  in the $J_{eff}=3/2$ space into the observable multipolar moments that are defined in the physical $J=9/2$ space of Np 5$f^3$ GS multiplet (see Methods). The calculated physical multipole order of NpO$_2$
is displayed in Fig.~\ref{fig_2}. Notice, that observable moments up to $K=$7 can exist  on an $f$-electron shell\cite{Santini2009}; we show the largest primary (odd) and secondary (even) order parameters as well as the physically important quadrupole order.  All non-zero multipole moments are listed in Supplementary Table~IV.  The physical dipole magnetic moments are found to completely vanish, since their contribution into the $\langle O_x\rangle$ and $\langle O_y\rangle$ pseudo-dipole $J_{eff}=3/2$ moments  is exactly canceled by that due to the $\langle T_x\rangle$ and $\langle T_y\rangle$  pseudo-octupoles. The primary order parameter is of rank-5
(triakontadipole); the physical octupole moments is an order of magnitude smaller and the magnitude of rank-7 moments is about 1/3 of that for triakontadipoles, in agreement with previous estimates for the relative contribution of those multipoles into the MMO order in NpO$_2$\cite{136_Santini_Garreta2006}. The magnetic triakontadipoles
on different sublattices are oriented in four different directions (forming
mutual angles corresponding to the angles between the cube's main diagonals) thus structure similarly to the 3\vk-AFM
dipole order in UO$_2$, see Fig.~\ref{fig_2}a. The secondary
order is dominated by  hexadecapole (rank-4), Fig.~\ref{fig_2}b; the ordered quadrupole moments (Fig.~\ref{fig_2}c) are roughly twice smaller. The quardupole order is directly related to the  pseudo-quadrupole one, since the $\Gamma_5$ (or $t_{2g}$) pseudo-quarupoles directly map into the physical ones, apart from swapping $xy \leftrightarrow yz$ and $x^2-y^2 \leftrightarrow xz$. The resulting physical quadrupole order can be represented,  in the space of   $\Gamma_5$ quadrupoles  [$O_{yz}$,$O_{xz}$,$O_{xy}$],  by four directions [$\bar{1}$11], [1$\bar{1}$1], [11$\bar{1}$], and [111] for four inequivalent Np sublattices [0,0,0], [1/2,1/2,0], [1/2,0,1/2], and [0,1/2,1/2], respectively. These quadrupoles can be depicted as  $\langle O_{z^2}\rangle$ moments with the principal axes $z$ along the corresponding direction at each given site (Fig.~\ref{fig_2}c). One sees that the ordered quadrupoles on the four Np sites forming the tetrahedron around each oxygen can either have their principal axes directed  along the Np-O bonds towards the central O,  or form the same angle of 70.5\textdegree\   with respect to those bonds. In both cases the tetrahedron symmetry is preserved. The first case is realized for two O along one of the principal  cubic diagonals, while the second one is found for 6 others, resulting in the lowering of NpO$_2$ symmetry to $Pn\bar{3}m$ from  $Fm\bar{3}m$ without any distortion of the cubic structure. This longuitudinal 3\vk\  quadrupole order, previously proposed on the basis of  resonant x-ray scattering \cite{Paixao2002}, and subsequently confirmed by  the splitting of $^{17}$O-NMR spectra in ordered NpO$_2$\cite{127_tokunaga}, is thus predicted by our {\it ab initio} SE Hamiltonian (\ref{HSE}).

\subsection*{Exchange splitting and magnetic excitations}

Having obtained the MMO of NpO$_2$ we subsequently calculated its excitation spectra, which has been previously measured, in particular, by INS\cite{Amoretti_1992,Magnani2008}.

The MMO lifts the degeneracy of CF GS $\Gamma_8$ quartet, the resulting exchange splitting calculated from the {\it ab initio} SE Hamiltionian (\ref{HSE}) is depicted
on the right panel of Fig.~\ref{fig1}a. The ground state is a $\Gamma_5$ singlet with the first excited doublet 
 found at 6.1 meV above the GS $\Gamma_5$ singlet and  the second excited level, singlet, located at 12.2 meV. The calculated position of first excited doublet  is in excellent agreement with the location of a prominent peak in INS spectra at about 6.4 meV\cite{Amoretti_1992} in the ordered phase; another  broad excitation was observed  in the range of 11-18 meV \cite{Magnani2008}, see below. 
 %, that is also in agreement with our calculated splitting. 
 Previously, an exchange  splitting of the $\Gamma_8$ 
 %that is very similar to our result on Fig.~ 
 was obtained assuming a diagonal uniform SEI between $\Gamma_5$ triakontadipoles tuned to reproduce the experimental position of the excited doublet\cite{136_Santini_Garreta2006}, thus obtained energy for the excited singlet is in agreement with our {\it ab initio} result.  %As shown in the same work of Santini {\it et al.}~\onlinecite{136_Santini_Garreta2006} , a primary  octupole $\Gamma_5$ MMO induces a heavy admixture of the excited $\Gamma_8$ quartet  into the GS, in disagreement with  specific heat measurement. Indeed, we obtain a rather moderate admixture of  the excited quartet into the ground-state $\Gamma_5$ singlet, just of about 9\%, due to the primary trikontadipole  order parameter predicted by our calculations. 
 
\begin{figure*}[!tb]
 	\begin{centering}
 		\includegraphics[width=1.9\columnwidth]{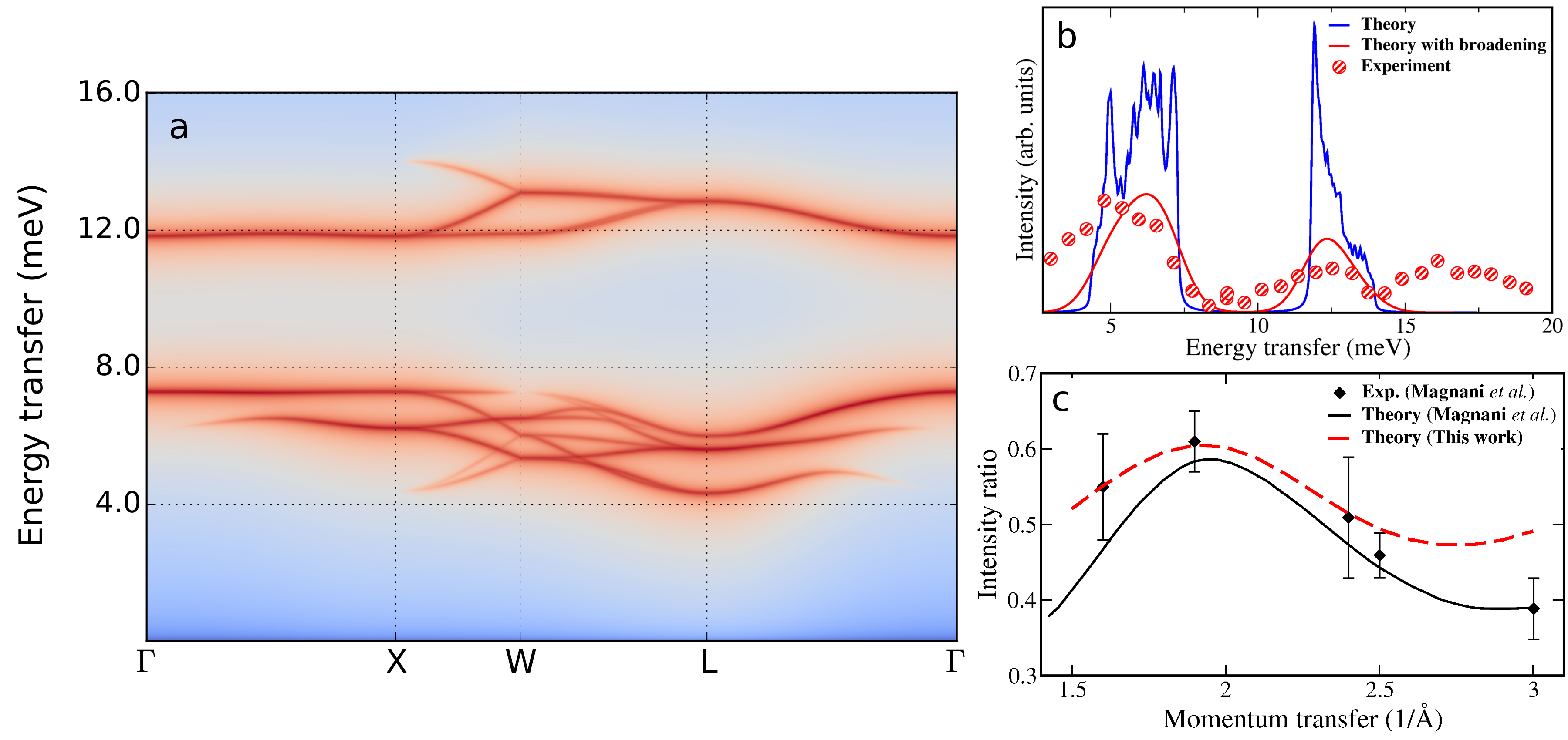} 
 		\par\end{centering}
 	\caption{{\bf Inelastic neutron-scattering (INS) spectra in ordered NpO$_2$.} {\bf (a)}  INS cross-section $S(\vq,E)$ with the momentum transfer $\vq$ along a high-symmetry path in the fcc Brillouin zone. The special points are $\Gamma=$[0,0,0], X=[1,0,0], W=$[1,\frac{1}{2},0]$, L=$[\frac{1}{2},\frac{1}{2},\frac{1}{2}]$, in units of $2\pi$/a. {\bf (b)}  Powder (spherically averaged) INS cross-section vs. energy transfer $E$ for $|q|=$2.5~\AA$^{-1}$. The  theoretical spectra was broadened with the Gaussian resolution function of 1.5~meV. The experimental points are from Magnani {\it et al.}~\cite{Magnani2008}   {\bf (c)} Ratio of the  spectral weights of the low-energy (around 6~meV) and high-energy (10-20~meV) features vs. momentum transfer in the powder INS spectra (panel b). The experimental points are from Magnani {\it et al.}~\cite{Magnani2008} Theoretical points from the same work are calculated with a semi-empirical SE Hamiltonian, see text.
 	}	
 	\label{fig3} 
 \end{figure*}

 We also calculated the theoretical INS cross-section, 
 $\frac{d^2\sigma(\vq,E)}{dEd\Omega}=S(\vq,E)\propto \sum_{\alpha,\beta}\left(\delta_{\alpha\beta}-\frac{q_{\alpha}q_{\beta}}{q^2}\right)\mathrm{Im} \chi_{\alpha\beta}(\vq,E)$, where $E$ ($\vq$)  the energy (momentum) transfer, respectively, $\alpha(\beta)=x,y,z$, and $\chi_{\alpha\beta}(\vq,E)$ is the dynamical magnetic  susceptibility. The latter was evaluated within the random-phase approximation (RPA)  from the full calculated {\it ab initio} SE Hamiltonian (\ref{HSE}), see Methods. The resulting INS spectra  $S(\vq,E)$ for $\vq$ along high-symmetry directions of the  fcc lattice is displayed in Fig.~\ref{fig3}a. Along the $\Gamma-X$ path it is similar to that previously calculated from the simplified empirical SE Hamiltonian of Ref.~\onlinecite{136_Santini_Garreta2006}. The $S(\vq,E)$ structure is richer along other directions showing multiple branches in the $E$ range from 4 to 8 meV. The experimental INS spectra $S(\vq,E)$ has not been measured so far due to lack of large single-crystal samples\cite{Santini2009}, hence, our result represents a prediction for future experiments. The calculated  spherically averaged INS cross-section $\int S(\vq,E) d\hat{q}$ is compared in Fig.~\ref{fig3}b with that measured\cite{Magnani2008} on powder samples at the same $|q|=$2.5~\AA$^{-1}$. The theoretical INS spectra exhibits prominent peaks at around 6~meV and at  about 12.5~meV, corresponding to the transition from the $\Gamma_5$ GS to the excited doublet and singlet, respectively. 
 
 The theoretical low-energy feature agrees very well with the measured INS spectra, once  experimental broadening is taken into account, see Fig.~\ref{fig3}b. The high-energy feature, though, is clearly split in experiment into two broad peaks centered at about 12 and 16~meV, respectively. In order to understand whether the relative weights of the low and high energy features is captured in the theory we employed the same analysis as  Magnani {\it et al.}~\cite{Magnani2008}. Namely, we evaluated, as a function of the momentum transfer, the ratio of  high-energy feature spectral weight to that of the low-energy one. The calculated ratio is in an excellent agreement with experiment up to  $|q|=$2.5~\AA$^{-1}$. As noticed in Ref.~\onlinecite{Magnani2008},  a phonon contribution to INS appears below 18~meV for large  $|q|$ thus rendering the separation of magnetic and phonon scattering less reliable for $|q| > $2~\AA$^{-1}$. The splitting of high-energy peak was clearly observed at all measured $|q|$; it was not reproduced by the simplified semi-empirical SEI employed by Magnani {\it et al.} They speculated that this splitting might stem from complex realistic Np-Np SEIs, which they could not determine from experiment. In the present work we determined the full SE Hamiltonian for the GS CF $\Gamma_8$ quadruplet. Hence, the fact that the splitting of INS high-energy feature is still not reproduced points out to its origin likely being a SE coupling between the ground-state and first excited  $\Gamma_8$ quaduplets (a significant contribution of the very high-energy $\Gamma_6$ CF doublet is unlikely).  This inter-quadruplet coupling can be in principle derived using the present framework; we have not attempted to do this in the present work. 
 
 Alternatively, lattice mediated interactions might be also considered as the origin of the high-peak splitting. Those interactions couple time-even moments, i.~e., the  quadrupoles within the $J_{eff}$ space. However, the QQ coupling is rather expected to impact the shape of low-energy peak in Fig.~\ref{fig3}b, since the corresponding lowest on-site excitation in the ordered phase (inset in Fig.~\ref{fig1}a)  is due to reverting of the on-site quarupole moment\cite{136_Santini_Garreta2006}. Indeed, we recalculated the theoretical INS spectra (Fig.~\ref{fig3}b) with  the magnitude of QQ block in the SEI (Fig.~\ref{fig1}b) scaled by a factor from 0 to 5. These variations of the QQ coupling strength do modify the shape of  low-energy peak but have no impact on the high-energy one. Since the lattice-mediated coupling is expected to modify exclusively the QQ block, it is thus quite unlikely to be the origin of the splitting.

\subsection*{Multipolar exchange striction}

The onset of the ``hidden order''   phase NpO$_2$ is marked by an anomalous volume contraction 
vs. decreasing temperature (anti-Invar anomaly). The  estimated total
volume contraction in the ordered state as compared
to the paramagnetic phase is  0.018\% at zero temperature\cite{Lander_nodistrotionNpO2}. 
This effect cannot be attributed
to the conventional volume magnetostriction, since ordered magnetic
moments are absent in NpO$_2$.  Hence, this anomaly was speculated\cite{2003JPCM_magnetovolume} to be induced by a  (secondary) quadrupole order coupling to the lattice. As we show below, this is not the case, and the volume contraction in  NpO$_2$ rather stems from the volume-dependent SEI coupling high-rank magnetic multipoles.
%The one-site magnetostriction
%due to magneto-elastic coupling of the quadrupolar moments to the
%lattice also appears to be very small since quadrupoles has very small
%value in the ordered state. 

\begin{figure}[!t]
	\begin{centering}
		\includegraphics[width=1.0\columnwidth]{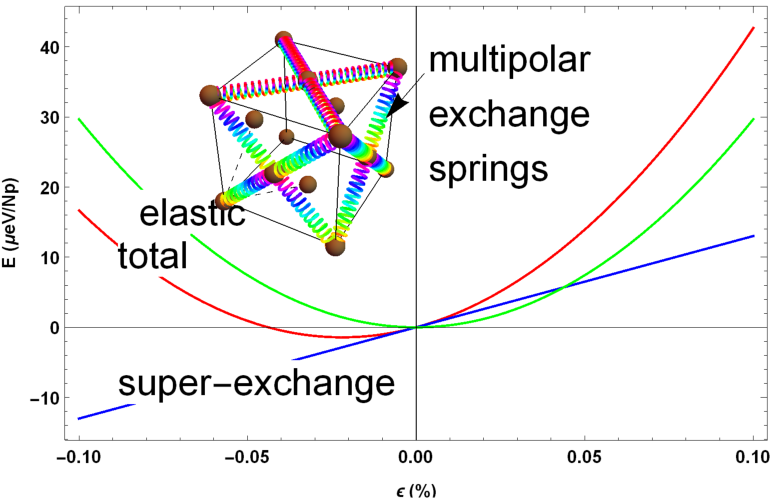} 
		\par\end{centering}
	\caption{{\bf Multipolar exchange striction in NpO$_2$.} The curves represent the  
		DFT+U elastic energy (green), ordering energy of the 
		multipolar ground state vs. volume (blue), the latter is calculated from the volume-dependent {\it ab initio} super-exchange interactions. The multipolar order
		is seen to induce the contraction of the equlibrium volume (total energy, red curve) due to the two-site multipolar
		striction effect, analogiously to  the  two-site volume magnetostriction
		in conventional magnetically ordered materials. 
		%Note: that even for
		%magnetically ordered materials the two-site volume magnetostriction
		%effect usually masked by one-site magnetostriction effect due to ``slave''
		%quadrupolar order. 
		The ``springs'' in inset schematically illustrate the action
		of the intersite super-exchange energy upon the  onset of NpO$_2$ multipolar order.}
	\label{fig4} 
\end{figure}

In order to make a quantitative estimation
for this anomaly we evaluated  the  volume dependence of NpO$_2$ ordering and elastic energies.  To that end we adopted the elastic constants calculated for NpO$_2$ in the framework
of DFT+U+SOC approach\cite{Wang2010}: $C_{11}=$404~GPa, $C_{12}=$
143~GPa. The corresponding parabolic volume dependence of elastic energy $E_{elast}=
1/3\left(C_{11}/2+ C_{12}\right)\epsilon^{2}\equiv K_{el} \epsilon^{2}$ where
$\epsilon=(V-V_0)/V_0$ is the volume contraction, $V_0$ is the NpO$_2$ 
%experimental 
 equilibrium volume\cite{Wang2010}, is depicted in Fig.~\ref{fig4}. The dependence of MMO energy vs. volume was obtained by calculating the SEIs  at a few different volumes and  then evaluating the mean-field order  and ordering energy vs. volume expansion or contraction. 
The SE ordering energy remains linear vs. $\epsilon$ in a rather large range ($\epsilon=\pm$1\% ); its dependence upon $\epsilon$ for the relevant range of  small $\epsilon$  is thus easily obtained.

 As is seen in Fig.~\ref{fig4}, the SE contribution shifts the equilibrium volume in the ordered state towards smaller volumes.  The negative slope for the ordering energy vs. volume is expected as the SEIs become larger with decreasing Np-Np distance.   Thus the multipolar
SEIs act as springs (see scheme in Fig.~\ref{fig4} inset) pulling Np atoms closer as the order parameters increase below $T_0$. Our approach is thus able to qualitatively capture this very small in magnitude subtle  effect: the calculated spontaneous multipolar exchange striction 
is 0.023 \% (Fig.~\ref{fig4}) at zero $T$ as compared to experimental estimate of 0.018\% \cite{Lander_nodistrotionNpO2,Caciuffo_JofPhys2003}. %The anomalous volume contraction of NpO$_2$ is thus quantitatively accounted for by the two-site multipolar exchange striction. 

 We also performed the same calculations suppressing the quadrupole-quadrupole SE obtaining only a minor change, by about 5\%, in the slope of MMO energy vs. volume. Hence,  the secondary quadrupole order plays virtually no role in the anomalous volume contraction. The physical origin of this effect is the volume dependence of leading, time-odd SEIs. 

To analyze the temperature dependence of the anomalous contraction, one may recast the linear-in-volume MMO energy (Fig.~\ref{fig4})  into a general form of $K_{SEI}^{pr}\epsilon \xi_{pr}^2(T)+K_{SEI}^{sec}\epsilon \xi_{sec}^2(T)$, where $\xi_{pr}(T)$ and $\xi_{sec}(T)$ are primary and  secondary order parameters, respectively, $K_{SEI}^{pr}$ and $K_{SEI}^{sec}$ are the  slopes of volume dependence for the corresponding contributions to MMO energy. 
 The temperature dependence of the anomalous volume contraction is thus given by that  of squares of the order parameters, $\xi_{pr}^2(T)$ and $\xi_{sec}^2(T)$.
%the temperature dependence of both SE and electron-lattice\cite{Santini2009} contribution to the  MMO energy are quadratic in the corresponding order parameters; 
As shown in Supplementary Fig.~2,  secondary quadrupole $\xi_{sec}^2(T)$ exhibits a smooth evolution across $T_0$ and rather slowly increase vs. decreasing $T$, while  primary time-odd $\xi_{pr}^2(T)$  features a discontinuity in the slope at $T_0$, as expected, with a rapid growth for $T<T_0$ , reaching about 60\% of total magnitude at $3/4 T_0$. This behavior of $\xi_{pr}^2(T)$  is in a perfect agreement with the shape of temperature dependence of the volume anomaly\cite{Lander_nodistrotionNpO2}, thus confirming that it is induced directly by the time-odd primary order.

\section*{Discussion}

In conclusion,  we have applied an advanced  {\it ab initio} framework to  the ``hidden-order" phase of Neptunium dioxide NpO$_2$. Our framework is based on the density-functional+dynamical mean-field theory (DFT+DMFT) in conjunction with a quasi-atomic approximation to local correlations on Np 5$f$. Its crucial part is a force-theorem method \cite{Pourovskii2016} that we employ to calculate super-exchange interactions between all multipole moments of the Np $f^3$ lowest crystal-field (CF)  manifold. From the resulting super-exchange Hamiltonian we derive  all order parameters of the ``hidden-order" phase, its ordering temperature, magnetic excitations and volume effect. In fact, numerous properties of the NpO$_2$``hidden-order" phenomenon that have been painstakingly determined in experiments over several decades --  absence of conventional magnetic order, the CF level scheme, the primary triandicapole order and secondary longitudinal 3\vk\ quadrupole order, the singlet-doublet-singlet exchange splitting of the CF ground state, the two-peak structure of the  magnetic-excitation spectra -- all of them are reproduced by our calculations that contain essentially no adjustable parameters.  Therefore, the present scheme is shown to provide full, parameter-free and  quantitatively correct description of super-exchange in complex realistic correlated insulators.  On the basis of our theory we may also identify the features that are not stemming from the inter-site coupling between the ground-state CF levels. For example, the splitting of the high-energy peak  in NpO$_2$  inelastic neutron scattering spectra likely stems from a super-exchange interaction with excited CF levels.  We also uncover an unconventional mechanism for the anomalous volume contraction observed in ordered NpO$_2$; which is induced not by the secondary quadrupole order coupling to the lattice as previously assumed, but rather by the volume dependence of leading time-odd super-exchange interactions.

This first-principles methodology -- a dynamical mean-field treatment of symmetric  paramagnetic phase combined with the force-theorem extraction of the full complex inter-site exchange responsible for the spontaneous symmetry breaking -- can be applied to a wide range of rare-earth, actinide and  heavy transition-metals correlated systems\cite{Santini2009,OppeneerRMP,Cameron2016,Witczak-Krempa2014}, in which the interplay of a large spin-orbit coupling with crystalline environment gives rise to a large degeneracy of the crystal-field ground state and high-rank multipole moments. ``Hidden" orders, stemming from coupling between those moments, can be predicted and their interplay with various parameters  -- external or chemical pressure, applied field,   lattice distortions -- identified, thus opening  up an avenue for theoretical search of new exotic phases of matter.

\section*{Methods}

\noindent 
{\bf \small DFT+HI first-principles calculations.} 
Our charge self-consistent DFT+DMFT calculations using the Hubbard-I (HI) approximation for Np 5$f$, abbriviated as DFT+HI, were carried out for the  CaF$_2$-type cubic structure of NpO$_2$ with the experimental lattice parameter $a=5.434$~\AA. We employed the Wien-2k full-potential code\cite{Wien2k} in conjunction with "TRIQS" library implementations for the DMFT cycle\cite{Aichhorn2016,Parcollet2015} and  HI. The spin-orbit coupling was included in Wien2k within  the standard second-variation treatment. The Brillouin zone (BZ) integration was carried out using 1000~{\bf k}-points in the full BZ and the local density approximation (LDA) was employed as DFT exchange-correlation potential.  

The Wannier orbitals representing Np 5$f$  states were constructed by the projective technique of Refs.~\onlinecite{Amadon2008,Aichhorn2009} using the Kohn-Sham bands enclosed by the energy window $[-2.04:2.18]$~eV around the Kohn-Sham Fermi energy; this window thus encloses all Np 5$f$-like bands. 
The use of a narrow window enclosing only the target (5$f$-like) band for the construction of  local orbitals results in a so-called ``extended" Wannier basis. By employing such a basis within DFT+HI one may effectively include  the contribution of hybdirization to the crystal-field splitting on localized shells,  as discussed in ref~\cite{Pourovskii2020}. The same choice for the local 5$f$ basis was  employed in our previous DFT+HI study\cite{Pourovskii2019} of UO$_2$, resulting in a good quantitative agreement of the  calculated CF splitting with experiment.

The on-site Coulomb interaction between Np 5$f$ was specified by the Slater parameter $F_0=$4.5~eV and the Hund's rule coupling $J_H$=0.6~eV; the same values were previously employed for UO$_2$\cite{Pourovskii2019}. The double-counting correction was computed using the fully localized limit (FLL)\cite{Czyzyk1994} with the atomic occupancy of Np $f^3$ shell \cite{Pourovskii2007}. The DFT+DMFT charge self-consistency was implemented as described in Ref.~\onlinecite{Aichhorn2011}. In our self-consistent DFT+HI calculations, we employed the spherical averaging of the Np 5$f$ charge density, following the approach of Delange {\it et al.}~\cite{Delange2017}, in order to suppress the contribution of LDA self-interaction error to the crystal field. The DFT+HI calculations were converged to 5 $\mu$Ry in the total energy.

\vspace{0.5cm}
\noindent 
{\bf \small Crystal-field and superexchange interactions.}  The self-consistent DFT+HI calculations predict a crystal-field (CF) split $^4$I$_{9/2}$ atomic multiplet to have the lowest energy, in agreement with Hund's rules for an $f^3$ shell; the calculated spin-orbit coupling $\lambda=$0.27~eV.  The predicted in these calculations CF splitting of the $^4$I$_{9/2}$  multiplet is shown in Fig.~\ref{fig1}a and the corresponding CF wavefunctions are listed in
 Supplementary Table~I. The cubic CF parameters -- $A_4^0\langle r^4\rangle$=-152~meV,  $A_4^4\langle r^4\rangle=5A_4^0\langle r^4\rangle$, $A_6^0\langle r^6\rangle$=32.6~meV, and $A_6^4\langle r^6\rangle=21 A_6^0\langle r^6\rangle$ -- were extracted by fitting the converged DFT+HI one-electron 5$f$ level positions\cite{Delange2017}.

The states of the CF ground-state quadruplet $\Gamma_8$ were labeled by projection $M$ of the pseudo-anglular quantum number $J_{eff}=$3/2 as specified in Supplementary Table~I.
We subsequently employed the FT-HI method\cite{Pourovskii2016} to evaluate all  SEIs  between the $J_{eff}=3/2$ quadruplet for several first Np-Np coordination shells. Previously the FT-HI method has been applied to systems with conventional magnetic primary order\cite{Pourovskii2019,Sunko2020}.  Within this method matrix elements of inter-site coupling $V^{\vR}$ for the Np-Np bond $\vR$ read:
\begin{equation}\label{V}
\langle M_1 M_3| V^{\vR}| M_2 M_4\rangle=\mathrm{Tr} \left[ G_{\vR}\frac{\delta\Sigma^{at}_{\vR_0+\vR}}{\delta \rho^{M_3M_4}_{\vR_0+\vR}} G_{-\vR}\frac{\delta\Sigma^{at}_{\vR_0}}{\delta \rho^{M_1M_2}_{\vR_0}}\right],
\end{equation}
where  $\rho^{M_iM_j}_{\vR_0}$ is the corresponding element of the $J_{eff}=3/2$ density matrix on site $\vR_0$, $\frac{\delta\Sigma^{at}_{\bf R_0}}{\delta \rho^{M_iM_j}_{\vR_0}}$ is the derivative of atomic (Hubbard-I) self-energy $\Sigma^{at}_{\bf R_0}$ over a fluctuation of the $\rho^{M_iM_j}_{\vR_0}$ element, $G_{\vR}$ is the inter-site Green's function for the Np-Np bond $\vR$ evaluated within the DFT+HI.  Once all matrix elements (\ref{V}) are calculated, they are transformed to the couplings $V_{KK'}^{QQ'}$ between on-site moments (\ref{eq:SE_bond}) as follows:
\begin{align}
V^{QQ'}_{KK'}({\vR})=
\sum_{\substack{M_1M_2 \\ M_3M_4}} \langle M_1M_3| V^{\vR'}|  M_2M_4\rangle \\ \left[O_{Q}^{K}(J)\right]_{M_2M_1} \left[O_{Q'}^{K'}(J)\right]_{M_4M_3}, \nonumber
\end{align}
where $\left[O_{Q}^{K}(J)\right]_{M_1M_2}$ is the $M_1M_2$ matrix element of the real spherical tensor defined in accordance with eq.~10 in Santini {\it et al.}~\cite{Santini2009}. The SEI matrix $\hat{V}$ shown in Fig.~\ref{fig1}b was subsequently obtained by rotating calculated $\hat{V}$ by 45\textdegree\   about the [010] axis thus aligning one of the NN bonds $\vR$ with $z$.

\vspace{0.5cm}
\noindent 
{\bf \small Mean-field calculations and analysis of order parameters.}  We solved the obtained SE Hamiltonian (\ref{HSE}) using the numerical mean-field package MCPHASE\cite{Rotter2004} including all 1$\vk$ structures up to 4$\times$4$\times$4 unit cells. We have also verified the numerical solution by an analytical approach, Namely, the mean-field equations read
\beq\label{eq:MF}
\langle \hat{O}_{a}\rangle=\frac{1}{Z}\mathrm{Tr}\left[\hat{O}_{a}\exp(-\beta H_{MF}^a)\right], 
\eeq
where $\hat{H}_{MF}^a=\sum_{b\ne a}\hat{O}_a\hat{V}_{ab}\langle \hat{O}_{b}\rangle$ is the mean-field Hamiltonian for sublattice $a$, $\hat{V}_{ab}$ is the SEI matrix between sublattices $a$ and $b$ ($\hat{V}_{ab}=\frac{1}{2}\sum_{[\vR_0+\vR]\in b} \hat{V}_{\vR}$ with $\vR_0 \in a$), $Z=\mathrm{Tr}\left[\exp(-\beta \hat{H}_{MF}^a)\right]$, $\beta=1/T$. Expanding the RHS of (\ref{eq:MF}) to the linear order in $\langle\hat{O}\rangle$ and  allowing for four inequivalent simple-cubic Np sublattices, which is a minimal number needed to cover
all possible ordered states on fcc lattice with a NN
coupling\cite{Smart_book},  one obtains a system of 60 linear equations. Solutions of the linearized MF equations
unambiguously  identify primary order parameters. With both numerical and analytical MF approaches  we obtained the  highest ordering temperature
(T=38K) and lowest free energy for the primary dipole-octupole
order displayed in Supplementary Fig.~1.

This order in the $J_{eff}=3/2$ space was subsequently  mapped into the physical $J=9/2$ space. 
The $J_{eff}$  density matrix on the inequivalent site $a$ reads    $\rho_a(J_{eff})=\hat{O}\cdot \langle \hat{O} \rangle_a$, where $\langle \hat{O} \rangle_a$ are the corresponding (pseudo) moments. This density matrix is upfolded into the one in the $J=9/2$ space, neglecting small contributions of the excited $J=11/2$ and $13/2$ multiplets and renormalizing the $\Gamma_8$ CF states accordingly, as $\rho_a(J)=|\Gamma_8\rangle\rho_a(J_{eff})\langle \Gamma_8|
$,
where $|\Gamma_8\rangle$ is the CF GS $\Gamma_8$ basis, with WFs written as columns in the order of  $M=-J_{eff}...J_{eff}$.  The physical $J=9/2$ moments are then calculated in the standard way,  $\langle O(J)_K^Q \rangle_a=\mathrm{Tr}\left[\rho_a(J)O(J)_K^Q\right]$.  

\vspace{0.5cm}
\noindent 
{\bf \small Calculations of dynamical susceptibility.}  In order to calculate the dynamical magnetic susceptibility  $\chi_{\alpha\beta}(\vq,E)$ we implemented a general RPA approach (see, e.~g., Ref.~\onlinecite{Rotter2012}).  Namely, the general susceptibility matrix in the $J_{eff}$ space reads
\beq
\bar{\chi}(\vq,E)=\left[I-\bar{\chi}_0(E)\bar{V}_{\vq}\right]^{-1}\bar{\chi}_0(E),
\eeq
where $\bar{\chi}_0(E)$ is the local bare susceptibility, $\bar{V}_{\vq}$ is the Fourier transform of SEI matrices $\hat{V}_{\vR}$, the bar $\bar{...}$ designate a matrix in combined $[a,\mu]$ indices, where $a$ label inequivalent  sublattices, $\mu=[K,Q]$ labels multipoles.  

The local susceptibility $\hat{\chi}_{0a}(E)$  for the  inequivalent site $a$ is calculated from the MF eigenvalues $E^a$ and eigenstates $\Psi^a$:

\beq
\chi_{0a}^{\mu\mu'}(E)=\sum_{AB}\frac{\langle \Psi_A^a|O_{\mu}|\Psi_B^a\rangle \langle \Psi_B^a|O_{\mu'}|\Psi_A^a\rangle}{E_B^a-E_A^a-E}\left[p_A^a-p_B^a\right],
\eeq
where  $A(B)$ labels four eigenvalues and eigenstates of the MF Hamiltonian  $\hat{H}_{MF}^a$ on the site $a$, $p_{A(B)}^a$ is the corresponding Boltzmann weight. 

Once the $J_{eff}$ susceptibility matrix $\bar{\chi}(\vq,E)$  is calculated, it is ``upfolded" to the physical $J=9/2$ space similarly to the $\rho_a(J_{eff})$ density matrix as described above.  The magnetic  susceptibility $\chi_{\alpha\beta}(\vq,E)$  is given by the dipole-dipole blocks  of the ``upfolded" $\bar{\chi}(\vq,E)$ summed over the sublattice indices.

\section*{Data availability}
The data that support the findings of this study are available from the corresponding author on reasonable request.

\bibliographystyle{naturemag_no_url} 
%\bibliography{bibliography}

\section*{ACKNOWLEDGMENTS} 

L.V.P. acknowledges  support by the European Research Council grants ERC-319286-"QMAC" and  the computer team at CPHT. S.K. is grateful to Ecole Polytechnique for financial support and to CPHT for
its hospitality.

\section*{AUTHOR CONTRIBUTIONS}
L.V.P. carried out the {\it ab initio} calculations and INS spectra simulations. S. K. carried out the order parameters' analysis. Both authors contributed equally to solving the mean-field equations, analysis of the results and writing of the manuscript.

\section*{COMPETING INTERESTS}
The authors declare no competing interests.

\end{document}